\documentclass[conference]{IEEEtran}
\IEEEoverridecommandlockouts
\usepackage{cite}
\usepackage{amsmath,amssymb,amsfonts}
\usepackage{graphicx}
\usepackage{xcolor}
\usepackage{multirow}
\usepackage{xcolor,colortbl}
\usepackage{subcaption} 
\usepackage{textcomp}
\usepackage{mwe}
\usepackage{algorithm}
\usepackage[noend]{algpseudocode}
\usepackage{diagbox}

\def\BibTeX{{\rm B\kern-.05em{\sc i\kern-.025em b}\kern-.08em
    T\kern-.1667em\lower.7ex\hbox{E}\kern-.125emX}}
    
\renewcommand\baselinestretch{0.915}

\begin{document}
\bstctlcite{IEEEexample:BSTcontrol}

\title{Runtime Deep Model Multiplexing for Reduced Latency and Energy Consumption Inference}
\author{
\IEEEauthorblockN{Amir Erfan Eshratifar}
\IEEEauthorblockA{\textit{Department of Electrical and Computer Engineering} \\
\textit{University of Southern California}\\
eshratif@usc.edu}
\and
\IEEEauthorblockN{Massoud Pedram}
\IEEEauthorblockA{\textit{Department of Electrical and Computer Engineering} \\
\textit{University of Southern California}\\
pedram@usc.edu}
\and
}

\maketitle

\begin{abstract}
We propose a learning algorithm to design a light-weight neural multiplexer that given the input and computational resource requirements, calls the model that will consume the minimum compute resources for a successful inference. Mobile devices can use the proposed algorithm to offload the hard inputs to the cloud while inferring the easy ones locally. Besides, in the large scale cloud-based intelligent applications, instead of replicating the most-accurate model, a range of small and large models can be multiplexed from depending on the input's complexity which will save the cloud's computational resources. The input complexity or hardness is determined by the number of models that can predict the correct label. For example, if no model can predict the label correctly, then the input is considered as the hardest. The proposed algorithm allows the mobile device to detect the inputs that can be processed locally and the ones that require a larger model and should be sent a cloud server. Therefore, the mobile user benefits from not only the local processing but also from an accurate model hosted on a cloud server. Our experimental results show that the proposed algorithm improves mobile's model accuracy by 8.52\% which is because of those inputs that are properly selected and offloaded to the cloud server. In addition, it saves the cloud providers' compute resources by a factor of 2.85$\times$ as small models are chosen for easier inputs.
\end{abstract}


\section{Introduction}
Deep learning is the rocket fuel of the recent advances in artificial intelligence and gaining popularity in intelligent mobile applications, solving complex problems like object recognition\cite{Resnet,decaf}, facial recognition\cite{BMVC2015_41,Sun:2014:DLF:2969033.2969049}, speech processing\cite{Amodei:2016:DSE:3045390.3045410}, and machine translation\cite{Bahdanau2014NeuralMT}. Although many of these tasks are important on mobile and embedded devices, especially for sensing and mission-critical applications such as health care and video surveillance, existing deep learning solutions often require powerful computational resources to run on. Running these models on mobile devices can lead to long run-times and the consumption of abundant amounts of resources, including CPU, memory, and power, even for simple tasks\cite{Canziani2017AnAO, Samragh2019AutoRankAR, abrishami}. Besides the enhancements achieved in optimizing the computation graph, efficient storage access such as Computational Storage Devices has shown promising results in further acceleration of deep learning models by reducing the data movements from storage device \cite{heydarigorji2020hypertune, heydarigorji2020stannis}.

\begin{figure}
\centering
\includegraphics[scale=0.5]{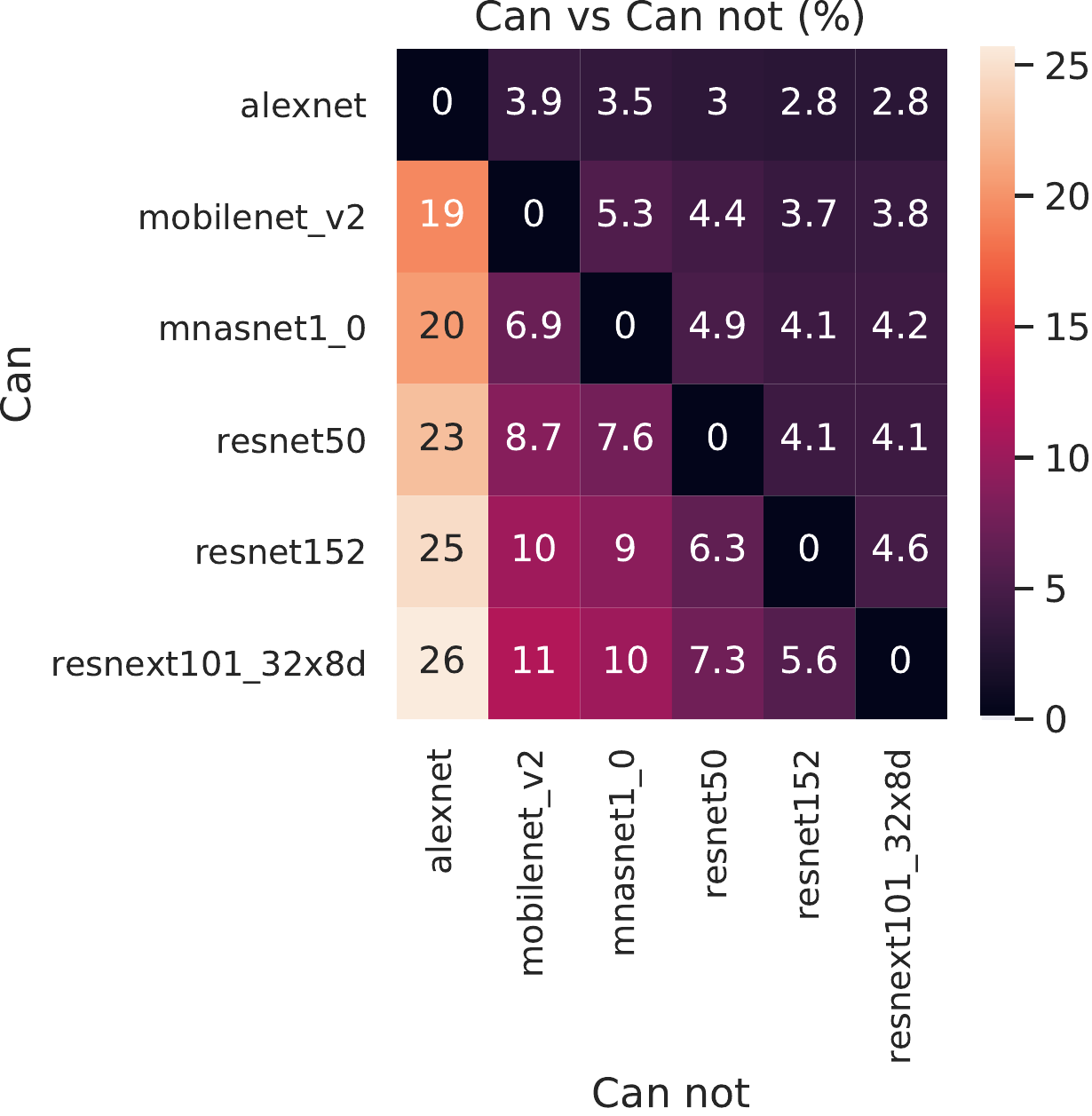}
\caption{The percentage of ImageNet's\cite{Deng2009ImageNetAL} validation set images that can be predicted correctly by a certain model but can not be correctly predicted by another model. As an example, \textit{alexnet}, as our worst performing model, can correctly predict 2.8\% of the inputs that the largest model, \textit{resnext101\_32x8d}, cannot. } \label{can_cant}
\end{figure}

The training process of deep neural networks (DNNs) is often offloaded to the cloud as it requires a huge amount of computations on large data. Once the model is trained, it will be used for inference on new unseen inputs. The inference process can be hosted privately on the local devices or as a public service in the cloud which we call \textit{mobile-only} and \textit{cloud-only} inference, respectively. In the cloud-only inference, the cloud providers grant access to the pre-trained models using an Application Programming Interface (API), which receives the input from the user and returns the inference results (predictions). The cloud-only inference is easy to deploy and scale up but compromises the data privacy and needs a reliable network connection. The communication cost of cloud-based inference can be also larger than the computation cost of running a small model locally. On the other hand, the mobile-only inference enables the mobile application to function without network access but is limited to small models due to the lack of enough computing resources. 

Recent promising advances in mobile-friendly deep architectures, such as \textit{mobilenet\_v2}\cite{Howard2017MobileNetsEC}, is closing the accuracy gap between the mobile and cloud level inference. For instance, the accuracy of \textit{mobilenet\_v2} as a mobile-scale model and \textit{resnext101\_32x8d} as a cloud-scale model are 73\% and 79\%, respectively. This essentially means that the mobile level model can predict 73\% of the inputs locally while the cloud level model can be called for the rest. As a result, a model multiplexer can be designed to call either the local model or the cloud model. However, the cost of this multiplexer should be kept small. We provide the definition of input complexity or easiness/hardness that we use throughout this paper:
\begin{itemize}
  \item Given a pair of small (mobile-side) and large (cloud-side) models, an input is easy if its label can be predicted correctly by the small model. An input is hard if the prediction is performed correctly by the large model.
  \item Given an ensemble of $N$ models, the complexity of an input lies in a range between 0 and $N$ representing the number of models that correctly predict the input's label. In the extreme case, the input complexity is 0, if all models can predict correctly. On the other hand, the input complexity is $N$ if no model can make a correct prediction on it.
\end{itemize}

\begin{figure*}[t]
\centering
\includegraphics[width=\linewidth]{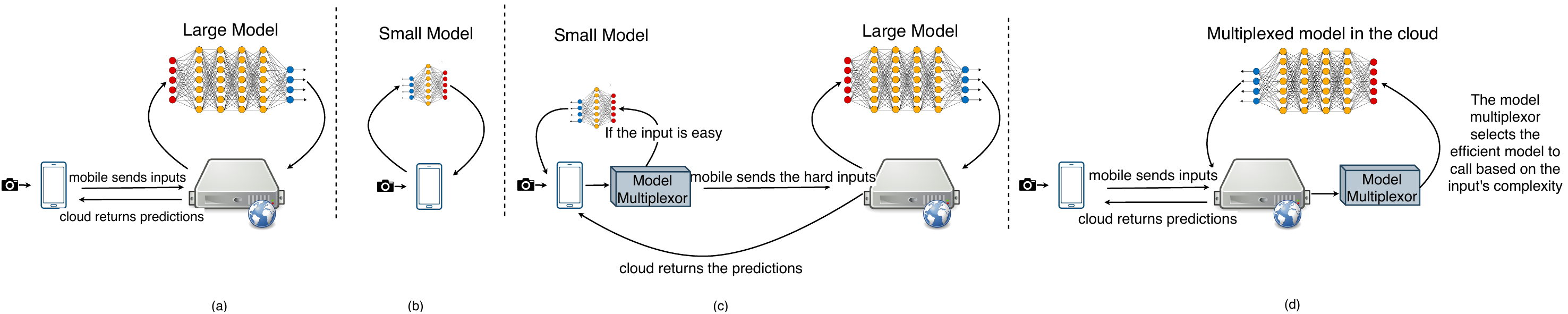}
\caption{Deep learning-powered mobile application deployment options. (a) and (b) show the status quo approaches of cloud-only and mobile-only approaches. In (c) a model multiplexer is called on the input which decides whether the input can be classified correctly on-device or should be offloaded to the cloud due to its complexity. (d) demonstrates multiplexing among a set of models (more than two) in the cloud intelligent service providers. } \label{mobile_or_cloud}
\end{figure*}

In the cloud inference services, the best-performing model is replicated across the servers and an API routes the users' input to one of the hosting servers. However, as we discussed earlier, a large portion of the inputs can be predicted correctly by worse-performing models with fewer computations. Also, a surprising fact is that the small model can predict some inputs correctly that the largest model cannot. For example, as demonstrated in Figure~\ref{can_cant}, the worst-performing model, \textit{alexnet}\cite{Krizhevsky:2017:ICD:3098997.3065386}, correctly predicts 2.8\% of the images that the best-performing model, \textit{resnext101\_32x8d}\cite{Xie2016AggregatedRT}, is not capable of. This suggests that if the multiplexing is performed well, the accuracy can be even higher than the most accurate model.

The proper selection of a model for inference can lead to considerable resource usage savings and higher accuracy. In this paper, we present a model multiplexer that receives the raw input (e.g. an image) and outputs a binary vector that shows the models capable of performing the inference. This multiplexer can be used in both mobile devices and cloud hosts. In a mobile application, the output of the multiplexer is a single binary value which decides whether the input should be processed locally or on the cloud. In a cloud service provider, instead of replicating the best performing models, we can host a wide range of different models on servers with different computing requirements and choose them depending on the complexity of the input. The multiplexer is a light-weight neural network extracting the required meta-features to speculate the correctness of the predictions of a set of models. We discuss the related works in the following.

Model compression techniques have been proposed to reduce the computational demand often by trading the prediction accuracy. These techniques include quantization \cite{ Han:2015:LBW:2969239.2969366, Rastegari2016XNORNetIC}, pruning \cite{ Han:2016:EEI:3007787.3001163}, optimized convolution operations \cite{ squeezenet,  Howard2017MobileNetsEC,Georgiev2017LowresourceMA}, and knowledge distillation for training small models using the knowledge of a teacher model \cite{Hinton2015DistillingTK}. Hardware-aware neural architecture search is also a recent interesting and promising research area \cite{Tan2018MnasNetPN}. These approaches require the user to be expert enough to come up with a specific model that satisfies the prediction accuracy requirements. Our proposed methods in this paper for model multiplexing enables the user to automatically select the model that requires the least resources.

Neurosurgeon~\cite{neurosurgeon} and JointDNN~\cite{8871124,10.1145/3194554.3194565} decides to offload some, or all layers in a DNN from the mobile device to the cloud server for reduced latency and mobile energy consumption. Unlike JointDNN, our granularity level is a complete DNN not a group of DNN layers. We seek to minimize the mobile inference latency by running the small models on the mobile side and large models on the cloud side depending on the hardness of input. Offloading the inference task to the cloud adds the additional cost of communication over a network which can be even larger than the computation cost. Besides, cloud-based inference compromises user privacy. However, if the mobile device can determine the input's complexity, it can run the inference locally as easy inputs can be solved by a small mobile-friendly DNN. Off-loading the DNN inference computations to the cloud can reduce the inference time\cite{Teerapittayanon17}, however, this is not always applicable because of privacy, communication latency, or connectivity issues. Another similar work\cite{Taylor:2018:ADL:3211332.3211336} uses hand-crafted features such as brightness or edge length in vision applications to choose the best model among a group of models which is highly dependent on the application domain. Furthermore, feature compression techniques are also proposed in prior arts to reduce the costs of uploading the inputs to the cloud server~\cite{8824955,8697647,8451100}.

Because the level of granularity in model multiplexing is a whole DNN, all acceleration techniques inside a DNN are complementary to our approach. Techniques such as convolutional kernel optimization\cite{Han:2016:EEI:3007787.3001163,Bhattacharya:2016:SSD:2994551.2994564}, task parallelism\cite{7460664}, and trading precision for time\cite{Huynh:2017:DMG:3081333.3081360} are used to accelerate the inference time to name but a few. Since a single DNN is not likely to meet all the constraints such as accuracy, latency, and energy consumption across inputs, a strategy to dynamically select the appropriate model to use appears to be a prudent option. 

Our approach is also related to ensemble learning where multiple models are used to solve an optimization problem. This technique is shown to be useful on many cognition tasks\cite{doi:10.1002/widm.1249}. However, in ensemble learning a voting mechanism (e.g. weighted mean) is used on all the models' predictions while our approach only calls a single model.

Figure \ref{mobile_or_cloud} illustrates the summary of four different scenarios that we addressed: (a) cloud-only inference where the input is always offloaded to the cloud, (b) mobile-only inference where the input is always processed locally, (c) mobile-cloud collaborative inference in which we choose between the mobile and cloud using the proposed multiplexer, (d) as the multiplexing can be done for more than two models, cloud API providers can also use the proposed algorithm to call smaller models instead of always calling the best-performing models. The paper makes the following contributions:
\begin{itemize}
  \item We present a deep learning-based approach to automatically learn how to multiplex DNN models depending on the input complexity and computational resource requirements. We leverage multiple DNN models and their expertise domain to improve the prediction accuracy and reduce the floating-point operations (FLOPs) and latency.
  \item The proposed method has a little overhead for the multiplexing as we use a small DNN acting as a pre-processor on the inputs. However, it benefits us by avoiding calling the expensive large models while achieving higher accuracy.
  \item In the mobile inference, the proposed method enables the mobile devices to perform the easy inference tasks locally and offload the hard ones to the cloud server. Therefore, it preserves the privacy of users for the inputs that are detected as easy.
  \item In the large scale cloud intelligent services, instead of replicating the best-performing model, one can host a range of small and large models and select from them at run-time depending on the input's complexity which will save the cloud resources by a factor of 2.85$\times$.
\end{itemize}

%


\section{Methodology}
In this section, we explain our proposed algorithm for model multiplexer design. Assume we are given $N$ models to multiplex from. We use a very light-weight mobile-friendly Convolutional Neural Network (CNN), consists of 4 convolutional layers, which outputs $N$ values in the range of [0,1]. The closer the $i$th value is to one, the more likely it is that the $i$th model can correctly predict the label. In this section, we explain our proposed method for learning the model multiplexer.

\begin{figure}
    \centering
    \begin{subfigure}[t]{\columnwidth}
        \raisebox{-\height}{\includegraphics[width=0.32\columnwidth]{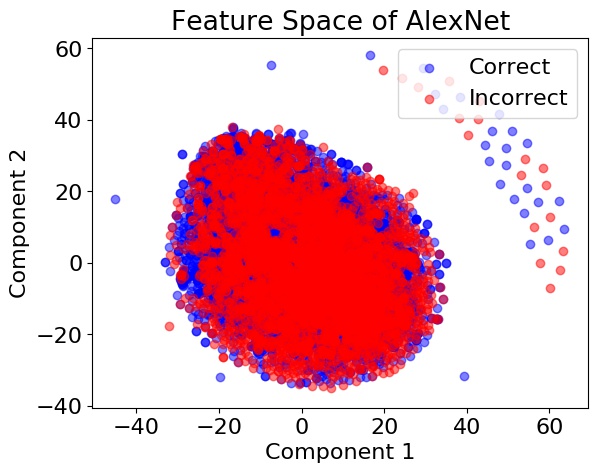}}
        \raisebox{-\height}{\includegraphics[width=0.32\columnwidth]{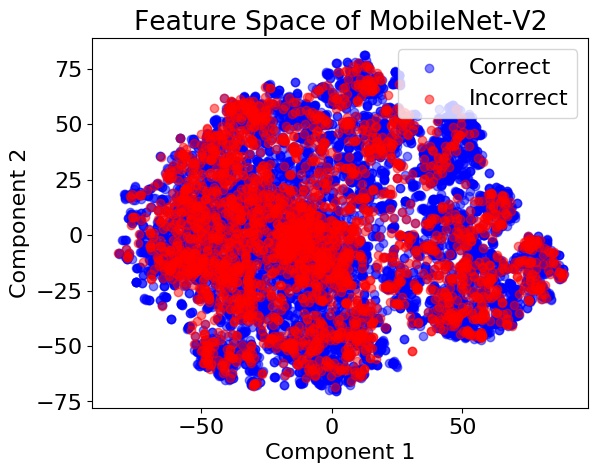}}
        \raisebox{-\height}{\includegraphics[width=0.32\columnwidth]{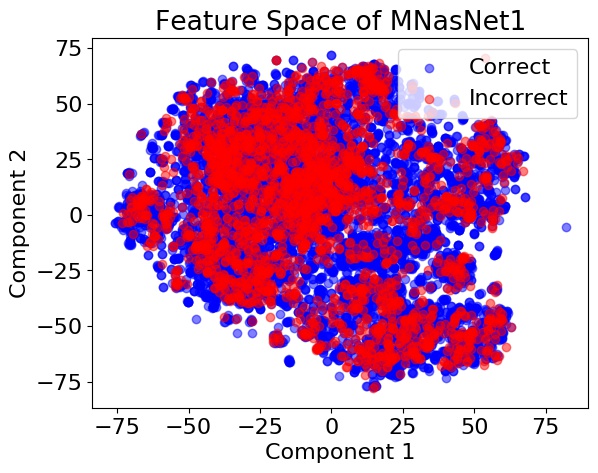}}
        \vspace{.6ex}
        \raisebox{-\height}{\includegraphics[width=0.32\columnwidth]{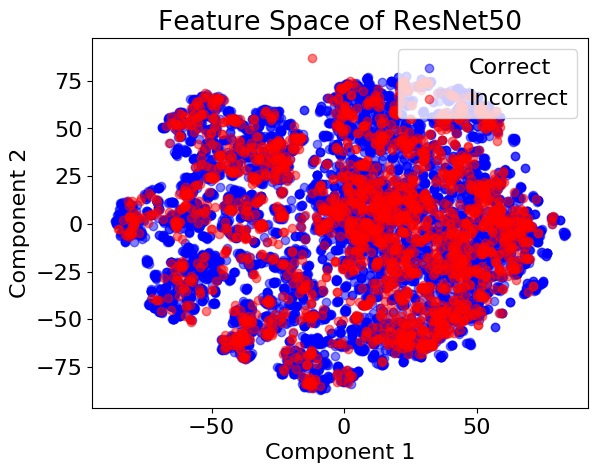}}
        \raisebox{-\height}{\includegraphics[width=0.32\columnwidth]{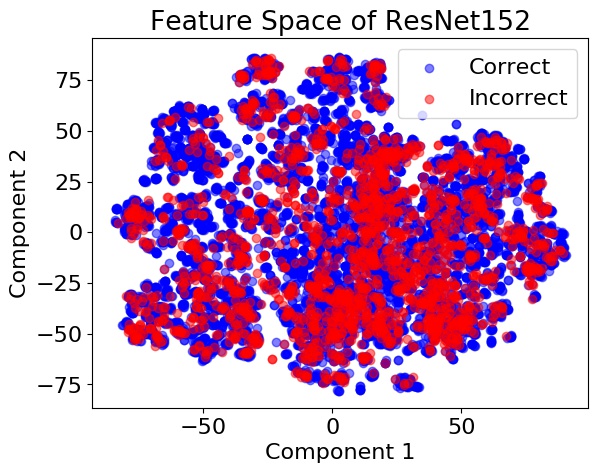}}
        \raisebox{-\height}{\includegraphics[width=0.32\columnwidth]{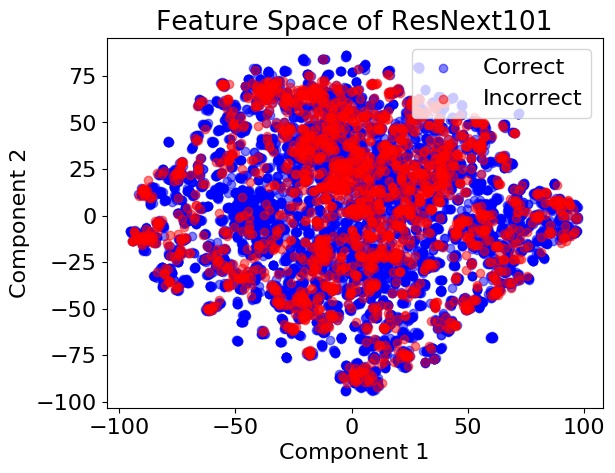}}
    \end{subfigure}
            \caption{The t-SNE visualization of feature space of our benchmark models on the validation set of ImageNet dataset. The feature space of correct and incorrect predictions are highly overlapped. This overlap shows that predicting whether the prediction of a certain model will be correct is a hard task. }
 \label{models_embeddings_cancant}
\end{figure}
\begin{figure}
\begin{center}
  \includegraphics[scale=0.15]{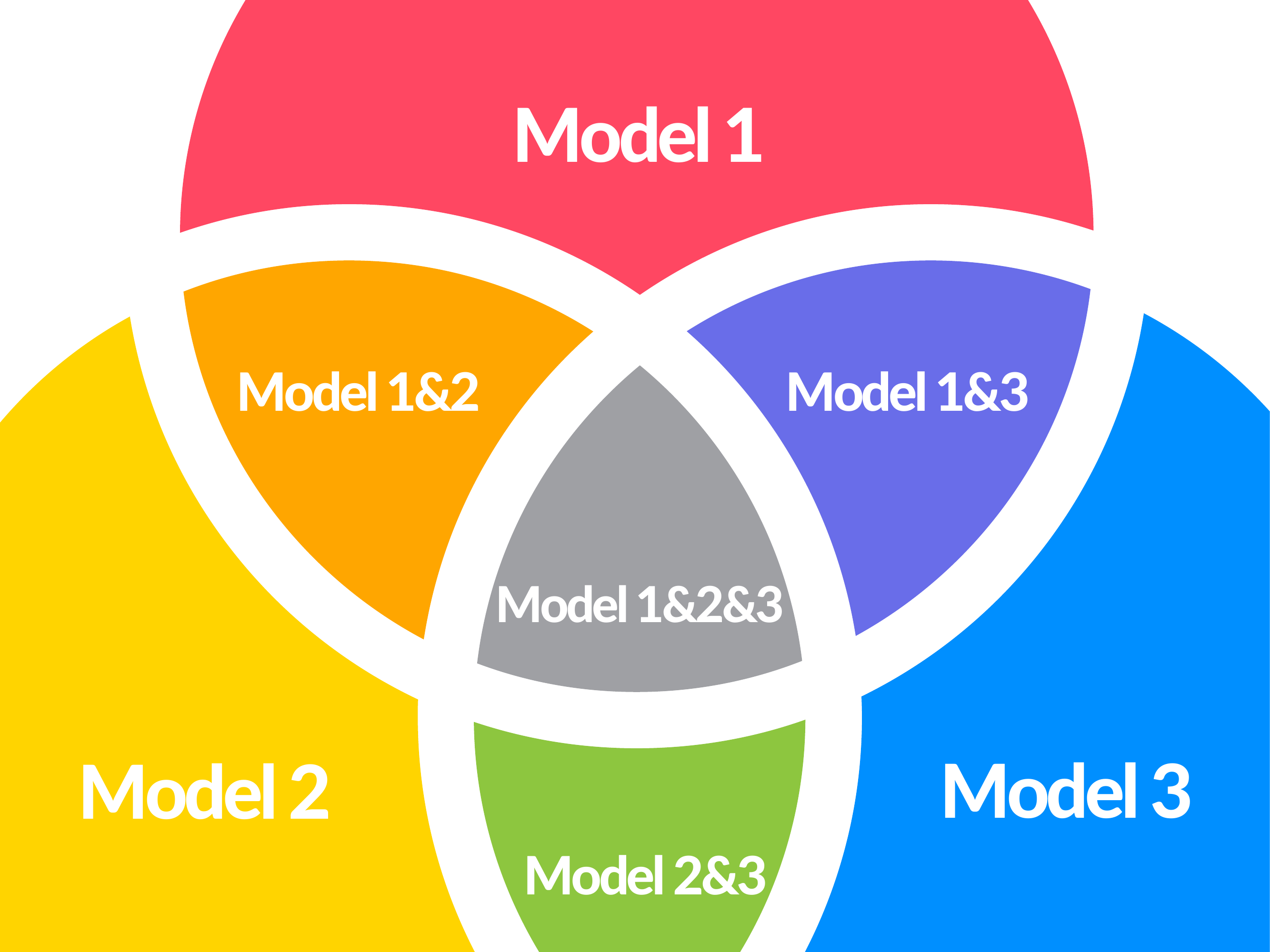}
  \caption{The target embedding space. The feature maps of the inputs are distributed in the space such that when a group of models can all predict the label of input correctly, their embeddings are close to each other. Also, when a group of models can predict the label correctly while another group of models can not, the distance between their embeddings is increased. This will lead to a feature space similar to a Venn diagram. For instance, the red region on top shows the samples which can be only predicted correctly by model 1.}
  \label{venn_loss}
\end{center}
\end{figure}

The output of the layer before the final classification layer in a deep neural network is a vector referred to as an \textit{embedding}. The embedding is the essential feature vector of the input learned by a neural network. Therefore, we expect the embeddings of different classes to shape in the space such that they are linearly separable. In Figure \ref{models_embeddings_cancant}, we have depicted the projected embeddings of the inputs which are predicted correctly or incorrectly by six different deep model benchmarks. The projection from the high dimensional space of embeddings into two-dimensional vectors is carried out using the t-SNE \cite{Maaten2008VisualizingDU} dimensionality reduction algorithm. Figure \ref{models_embeddings_cancant} shows that there is no separation between the inputs which are predicted correctly or incorrectly by a certain model. As a result, using a pre-trained deep model for the model multiplexing without any further supervision would be ineffective. We propose a loss function, referred to as \textit{contrastive loss}, for jointly training all the models we are multiplexing from. The intuition behind the contrastive loss is that given two groups of models if one group can predict the label of input correctly and the other group cannot, the distance between their embeddings will be increased. Also, when a group of models all can predict an input correctly, the distance between their embeddings will be decreased. This loss function shapes the embedding space of models similar to a Venn diagram. As depicted in Figure \ref{venn_loss}, for example, the red region on top contains the samples which can be predicted correctly only by Model 1 whereas the gray region in the center is the embedding space of samples which are predicted correctly by all models. The proposed loss is inspired by the Pairwise Ranking Loss\cite{Chen2009RankingMA} in which the distance of representations of the samples is determined by the pairwise similarity of the samples.
\begin{figure}
\begin{center}
  \includegraphics[width=\columnwidth]{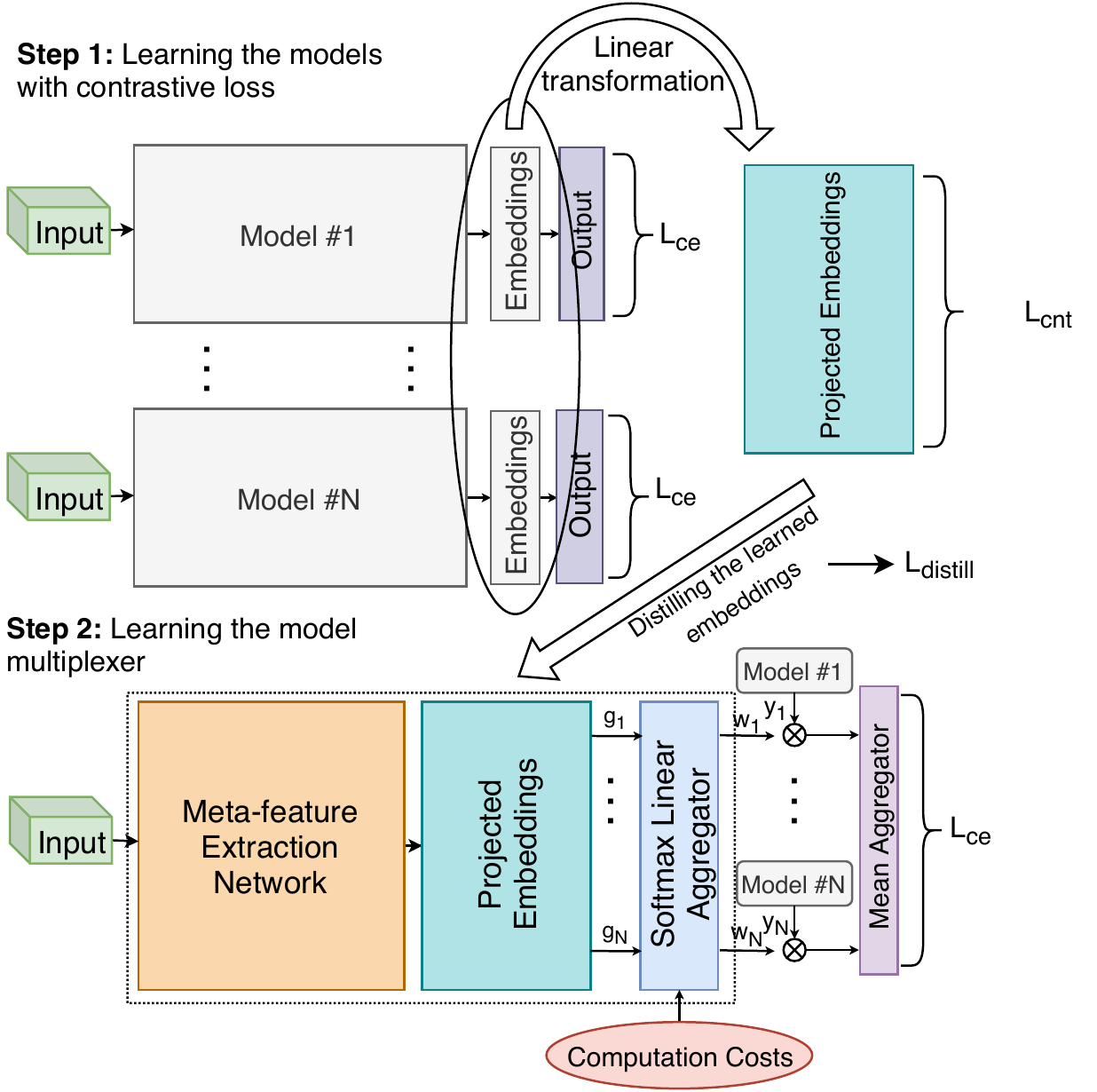}
  \caption{Model multiplexer training procedure and its architecture. In the first step, the models we are multiplexing from are trained using the contrastive loss. The contrastive loss allows the learned embeddings to be grouped into regions where each region determines the expertise domain of a subset of models. In the second step, we distill the learned embeddings from the first step into the multiplexer by adding a distillation loss function. The multiplexer outputs a set of weights where each weight determines the confidence of its corresponding model about the prediction correctness. We also show where each loss function is applied to in the figure.}
  \label{multiplexer_architecture}
\end{center}
\end{figure}

Once the models are trained using the contrastive loss, we need to train the model multiplexer using our trained models. As we discussed earlier, given $N$ models, the model multiplexer will have $N$ outputs where the $i$th output shows the probability that $i$th model can predict the input correctly. One advantage of using multiple models is that we can also leverage the ensemble techniques. In an ensemble model, a subset of models is selected for the inference and the mean of the selected models' outputs will be the final prediction. Our training procedure for model multiplexer allows for selecting more than one model for ensembling purposes so as to increase the accuracy. The training procedure of both CNN models using contrastive loss and model multiplexer will be discussed in the following sections. 

\subsection{Contrastive Loss Function}
We seek to learn the features which are useful for extracting the domain expertise of a group of models. By expertise, we specifically mean the set of inputs that can be predicted correctly by a certain model. In practice, since the embedding vector size of models can be different, we define $h_i$ which will linearly transform the embedding space of $i$th model into the same dimension and further normalize the linearly transformed embeddings by $L_2$ norm. We call this transformed space \textit{projected embeddings}. An embedding and a projected embedding of $i$th model are shown as $g_i$ and $e_i$, respectively:

\begin{equation} \label{eq:distance_function}
\begin{aligned}
e_i = normalize(h_i^T g_i)
\end{aligned}
\end{equation}

Given a pair of models, three cases can happen regarding their capability of correct prediction: 1- Both can predict correctly in which case we decrease the distance between the projected embedding vectors. 2- One can predict correctly whereas the other cannot in which case we increase the distance between the projected embedding vectors. 3- None of them can predict correctly in which we will not apply the contrastive loss and let the cross-entropy loss enable the models to learn the correct prediction without any interference from the contrastive loss. With this explanation, the contrastive loss function, $L_{cnt}$, will be of the form:

\begin{equation} \label{eq:contrastive_loss}
\begin{aligned}
L_{cnt}(\hat{y},y) = \sum_{i=1}^N \sum_{j=1; i\neq j}^N \log(d(e_i, e_j)) (&(\hat{y_i}==y\;\&\;\hat{y_j}==y)\;\\
    -&(\hat{y_i}!=y\;\&\;\hat{y_j}==y)\;\\
    -&(\hat{y_i}==y\;\&\;\hat{y_j}!=y))\;
\end{aligned}
\end{equation}
where $y$ is the true label, $\hat{y_i}$ is the prediction of $i$th model, $d$ is a distance function. We may choose $d$ as any family of functions satisfying $d: \{E_1, E_2\} \rightarrow [0,1]$, where $E_1$ and $E_2$ are embedding space domain. We use the cosine distance for the distance function as following:

\begin{equation} \label{eq:distance_function}
\begin{aligned}
d(e_1, e_2) = \frac{e_1^T e_2}{e_1^T e_1 * e_2^T e_2}
\end{aligned}
\end{equation}

Other distance functions in which the output range is normalized to $[0,1]$ can be used in this formulation, however, we performed the experiments using the cosine distance. We train all the models that we are multiplexing from by adding the contrastive loss to their main loss function which is cross-entropy in our case. Figure \ref{multiplexer_architecture}'s Step 1 demonstrates the learning procedure with the contrastive loss which is applied to all models in the ensemble.

\subsection{Learning the Model Multiplexer}
Let $f_i : X \,\to\, y, \forall i\in \{1,..,N\}$ denote the learned prediction functions of $N$ deep learning models, where $X$ and $y$ are the input space and target predictions, respectively. Similar to standard stacking\cite{Breiman1996}, we seek to determine a weighted prediction function of the form:

\begin{equation} \label{eq:standard_stacking}
y_{ENS} = \sum_{i=1}^N w_i(x) f_i(x), \forall{x\in X}
\end{equation}
where $w_i(x) \in \mathbb{R}$ is the $i$th model contribution to the final prediction. Let $m_i$ represent the meta-feature extraction function for predicting the correct prediction of $i$th model, and $c_i$ denote the computing cost of the $i$th model. The meta-features are supposed to learn the features necessary for determining the weights that corresponds to the likelihood that a certain model can make a correct prediction on the given input. We model $w_i(x)$ as a linear function of the meta-features weighted by the inverse of the computing cost which is FLOPs in our case:

\begin{equation} \label{eq:stacking_weights}
w_i = \sum_{j=1}^M \frac {v_{ij} m_j(x)} {c_i}, \forall{x\in X}
\end{equation}
where $v_{ij} \in \mathbb{R}$. To squash $w_i$ into the range of $[0,1]$, we normalize them using Softmax function. Under this assumptions, Equation~\ref{eq:standard_stacking} can be rewritten as:

\begin{equation} \label{eq:standard_stacking_rewritten}
y_{ENS} = \sum_{i=1}^M \frac{\exp(\sum_{j=1}^M \frac {v_{ij} m_j(x)} {c_i})} {\sum_{k=1}^M \exp(\sum_{j=1}^M \frac {v_{kj} m_j(x)} {c_k})} f_i(x), \forall{x\in X}
\end{equation}

We parameterize all $m_i$ with a convolutional neural network and denote its parameters by $\Theta$. As a result, the learnable parameters are $\Theta$ and $v_{ij}$. This formulation leads to the following optimization problem:

\begin{equation} \label{eq:optimization}
min_{\Theta, v} L_{mux}(y_{ENS}, y) = \sum_{x \in X} y(x) \log(y_{ENS}(x))
\end{equation}

where $X$ is the training set. We also add a distillation loss for distilling the projected embeddings of all models learnt by the contrastive loss into the multiplexer. We denote the projected embedding learnt by the $i$th model as $e_i$ and the $i$th meta-feature of the model multiplexer as $g$:

\begin{equation} \label{eq:distill_loss}
L_{distill}(g,e) = \sum_{i=1}^N d(m, e_i)
\end{equation}
where $d$ is the same function as in Equation \ref{eq:distance_function}.

\begin{algorithm}
\caption{Model multiplexer learning}\label{alg:model_multiplexer_learning}
\begin{algorithmic}[1]  
    \State {Initialize all $N$ models parameters, $\theta_i$}
    \State {Initialize the model multiplexer parameters, $\Theta$, $v$}
    \State //Learning the models we are multiplexing from.
     \For{\texttt{iteration = 1,2,...}}
        \State Sample a batch of inputs $x$ with labels $y$
        \For{all models $i$}
            \State $\hat{y_i}$ = $f_{\theta_i}(x)$
        \EndFor
        \For{all models $i$}
            \State $L_i = L_{cnt}(\hat{y}, y) + L_{ce}(\hat{y_i},y)$
            \State $\theta_i$ = $\theta_i$ - $\alpha \nabla L_{i}$
        \EndFor
      \EndFor
      \State //Learning the model multiplexer.
      \For{\texttt{iteration = 1,2,...}}
        \State Sample a batch of inputs $x$ with labels $y$
        \For{all models $i$}
            \State $\hat{y_i}, e_i$ = $f_{\theta_i}(x)$
            \State $\hat{w_i}, m$ = $f_{\Theta}(x)$
        \EndFor
        \State $y_{ENS} = \sum_{i=1}^N w_i(x) \hat{y_i}$
        \State $L = L_{mux}(\hat{y}, y) + \sum_{i=1}^N L_{distill}(m,e_i)$
        \State $\Theta$ = $\Theta$ - $\alpha \nabla L$
      \EndFor
    \State \Return $\Theta$
\end{algorithmic}
\end{algorithm}

Figure \ref{multiplexer_architecture}'s Step 2 demonstrates the proposed learning algorithm for training the model multiplexer. The complete training process of the models and the multiplexer is demonstrated in Algorithm \ref{alg:model_multiplexer_learning}.

\subsection{Multiplexing process}
\begin{algorithm}
\caption{Multiplexing process}\label{alg:multiplexing}
\begin{algorithmic}[1]
    \State Inputs: x is the model input, T is the weight threshold
    \State $w$ = $f_\Theta(x)$
    \State $S = argmax(w)$ or $S$ = NoneZeroElements($(w>T)$)
    \State $\hat{y}$ = $avg(f_s(x)), \forall{s\in S}$
    \State \Return $\hat{y}$
\end{algorithmic}
\end{algorithm}

We explained how to train the model multiplexer. The multiplexing can be performed in two ways: 1- We find the maximum weight and call the corresponding model to perform the inference. 2- We select all models whose corresponding weight is greater than a threshold and take the average of their outputs. The whole multiplexing process is shown in Algorithm \ref{alg:multiplexing}.

\section{Experiments and Results}
\subsection{Experimental Setup}
\textbf{Hardware.} We evaluate our approach on the NVIDIA Jetson TX2 embedded deep learning platform as our mobile device. The system has a 64 bit dual-core Denver2 and a 64 bit quad-core ARM CortexA57 running at 2.0 GHz, and a 256-core NVIDIA Pascal Graphics Processing Unit (GPU) running at 1.3 GHz. The board has 8 GB of LPDDR4 RAM and 96 GB of storage (32 GB eMMC plus 64 GB SD card). We use NVIDIA GTX 1080Ti as our server-side hosting GPU. We measure the energy consumption of each component on the board using the INA226 power sensor. We use and set to the average Wi-Fi uplink and downlink speed in the United States\cite{speedtest} for the communication latency.\\
\textbf{System Software.} Our evaluation platform runs Ubuntu 16.04 with Linux kernel v4.4.15. We use PyTorch\cite{pytorch}, cuDNN (v7.0) and CUDA (v10.1).\\
\textbf{Deep Learning Models.} We consider six of the state-of-the-art CNN models for image recognition. The models are built using PyTorch and trained on the ImageNet ILSVRC 2012\cite{Deng2009ImageNetAL} training set. The total number of floating-point operations required for a single inference is used as the computation cost of the model in Equation \ref{eq:stacking_weights}. We train all the benchmark models and the multiplexer model for 200 epochs on the training set of ImageNet.

\subsection{Results}

\begin{table}
\caption{The latency, percentage of local inference, and accuracy of mobile-only, cloud-only and hybrid (multiplexing) methods. \textit{mobilenet\_v2} and \textit{resnext101\_32x8d} are used as the mobile and cloud deep models, respectively.}
  \centering
\label{tab:result_mobile}
\begin{tabular}{|c|c|c|c|c|c|}
\hline
Setup & Flops & Latency & Mobile Energy & Local & Acc.\\
\hline
Mobile-only & 299M & 3.53ms & 12mJ  &   100\%  & 71.88\% \\
\hline
Cloud-only  & 16.4G & 13.1ms  & 110mJ &   0\%  & 79.39\% \\
\hline
\textbf{Hybrid} & \textbf{5.75G} & \textbf{10.12ms} & \textbf{55.36mJ}  &   \textbf{68\%}  & \textbf{80.4\%} \\
\hline
\end{tabular}
\end{table}

\textbf{Mobile-cloud collaborative inference.} In this scenario, one light-weight model is hosted on the mobile side (\textit{mobilenet\_v2}) and the best-performing model (\textit{resnext101\_32x8d}) on the cloud side. The multiplexer is a 4-layered light-weight CNN adding negligible computation cost compared to the mobile-hosted model. Our neural multiplexer outputs a single value between zero and one. Zero means the input should be classified on the mobile device and one means the input should be classified on the cloud server. We use a threshold function at 0.5 to binarize the output. We call the multiplexer to decide whether to perform the inference on the mobile devices or the cloud server. Although a negligible extra computation is added to the mobile inference, it benefits the user with about 10\% improvement in the accuracy which is because of the inputs which could be classified correctly only by the cloud's large and accurate model. In order to have a clear understanding of the components of the latency and energy consumption, we provide their formulations. The latency and energy consumption of a single inference using the mobile-only approach is only due to the computations required for the inference using the mobile-side model (\textit{mobilenet\_v2}). We refer to both latency and energy consumption as the cost which is represented by $C$:
\begin{equation} \label{eq:mobile_costs}
C_{mobile-only} = C_{mobile-compute-inference}
\end{equation}

The latency and energy consumption of a single inference using the cloud-only approach consists of the communication costs, and the cloud compute costs:
\begin{equation} \label{eq:cloud_costs}
\begin{aligned}
C_{cloud-only} &= C_{upload} + C_{cloud-compute-model} \\
               &+ C_{download}
\end{aligned}
\end{equation}

The latency and energy consumption of a single inference using the hybrid approach has two possible cases: 1- The multiplexer decides to perform the inference locally in which:
\begin{equation} \label{eq:hybrid_costs_1}
\begin{aligned}
C_{hybrid-m} &= C_{mobile-mux} + C_{mobile-compute-inference}
\end{aligned}
\end{equation}
2- The multiplexer decides to perform the inference on the cloud in which:
\begin{equation} \label{eq:hybrid_costs}
\begin{aligned}
C_{hybrid-c} &= C_{mobile-mux} + C_{upload}\\
           &+ C_{cloud-compute} + C_{download}
\end{aligned}
\end{equation}
Therefore, the cost of the hybrid approach will be the weighted average of the two previous equations. The weights are determined by the percentage of inferences that are performed on the mobile and cloud. The hybrid approach's cost will be:
\begin{equation} \label{eq:hybrid_costs}
\begin{aligned}
C_{hybrid} &= (\% local) * C_{hybrid-m} + (\% cloud) * C_{hybrid-c}\\
\end{aligned}
\end{equation}
Detailed results for the collaborative inference between the mobile device and cloud server are shown in Table \ref{tab:result_mobile}. As it shows, 68\% of the inputs are decided by the multiplexer to be processed locally on the mobile device while the other 32\% are offloaded to the cloud. Our algorithm also improves the accuracy of the mobile-only approach by 8.5\% which is because of the correct predictions on those inputs that are offloaded to the cloud. The accuracy of the hybrid approach is even higher than the cloud model which is because of the fact the small model can make correct predictions on inputs that the large model cannot. The True Negative Rate of the multiplexer is the detection rate of the inputs that can be classified correctly by the mobile device which is 0.966\% in our case. This means we miss (1-0.966)*0.7188=2.4\% of the inputs that could be predicted correctly by the mobile device which will be compensated by the powerful cloud model. The latency and energy of the hybrid approach in Table \ref{tab:result_mobile} is worse than those of the mobile-only but this comparison is not fair. The reason is that the extra latency and energy cost we pay is directing increasing the accuracy. Neglecting the cost of multiplexing, the extra latency, and energy is because of two reasons: 1- The inputs that could be predicted correctly on the mobile but we offload it to the cloud which is only the case for 2.4\% of the inputs; 2- The inputs that could not be predicted correctly on the mobile and we offload it to the cloud which is the case for 32-2.4\%=29.6\% of the inputs and is the dominant component.

\begin{table}
\caption{The FLOPs, latency, accuracy of six of the state-of-art CNN models. The \textit{Called} column shows the percentage of inputs which are decided to be predicted by the corresponding model.}
  \centering
\label{tab:result_cloud}
\begin{tabular}{|c|c|c|c|c|}
\hline
Model & FLOPs & Latency & Accuracy & Called \\
\hline
\textit{alexnet}\cite{Krizhevsky:2017:ICD:3098997.3065386} & 655M &  6.8ms &  56.55\%  & 10.56\% \\
\textit{mobilenet\_v2}\cite{Howard2017MobileNetsEC} & 299M &  3.0ms   &   71.88\%  & 18.80\% \\
\textit{mnasnet1\_0}\cite{Tan2018MnasNetPN} & 313M &  5.5ms  &   73.45\%  & 21.80\% \\
\textit{resnet50}\cite{Resnet} & 4.08G &  8.9ms &   76.15\%  & 14.80\% \\
\textit{resnet152}\cite{Resnet} & 11.5G &  11.3ms &   78.31\%  & 15.80\% \\
\textit{resnext101\_32x8d}\cite{Xie2016AggregatedRT} & 16.4G &  11.8ms &   79.31\%  & 18.24\% \\
\hline
\textbf{Hybrid-single} & \textbf{5.75G} &   \textbf{7.73ms}  & \textbf{83.86\%} & 100\% \\
\textbf{Hybrid-ensemble} & \textbf{7.12G} & \textbf{8.15ms} & \textbf{85.54\%} & 100\% \\
\hline
\end{tabular}
\end{table}

\textbf{Cloud-based API inference.} As in the cloud-hosted inference services the best-performing model is replicated on the servers while many inputs are easy and can be processed with small models. The proposed algorithms help to distribute the easy and hard inputs to the model that will consume minimum resources. Table \ref{tab:result_cloud} demonstrates the improvements we could achieve for the cloud providers. The \textit{hybrid-single} represents the scenario in which we multiplex a single model from a group of models while \textit{hybrid-ensemble} represent the scenario in which we multiplex more than one model from a group of models. The models whose associated weight in the Equation \ref{eq:standard_stacking_rewritten} is greater than a threshold are selected to perform the inference. We sweep over all possible values for the threshold and found 0.288 as the best value giving the maximum accuracy. Similarly, we also show the cost equation for the hybrid approach of cloud-based inference:
\begin{equation} \label{eq:hybrid_costs_cloud}
\begin{aligned}
C_{hybrid} &= \sum_i (\% called_i) * C_{i_{cloud-compute-model}}\\
\end{aligned}
\end{equation}
where $called_i$ is the percentage of the times that the $i$th model is called, and $C_{i_{cloud-compute-model}}$ represents the cost of running $i$th model on the cloud. In the \textit{hybrid-single} case, the FLOPs count is reduced from 16.4G (i.e. the largest model FLOPs) to 5.75G which essentially results in saving GPU resources by a factor of 2.85$\times$. The latency is reduced by 34.5\% and the over accuracy is improved by 4.55\%. In addition, if we use more than model after the multiplex. i.e. ensembling the models, we can further improve the accuracy. Although ensembling increases the FLOPs, we exploit the fact that model ensembles can be parallelized on GPUs. As a result, the increase (\%) in the latency of \textit{hybrid-ensemble} is less than the increase (\%) in its FLOPs.

\begin{figure}
    \centering
    \begin{subfigure}[t]{\columnwidth}
        \raisebox{-\height}{\includegraphics[width=0.49\columnwidth]{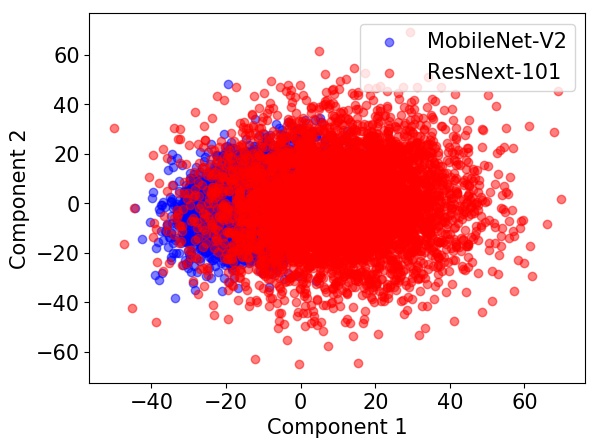}}
        \raisebox{-\height}{\includegraphics[width=0.49\columnwidth]{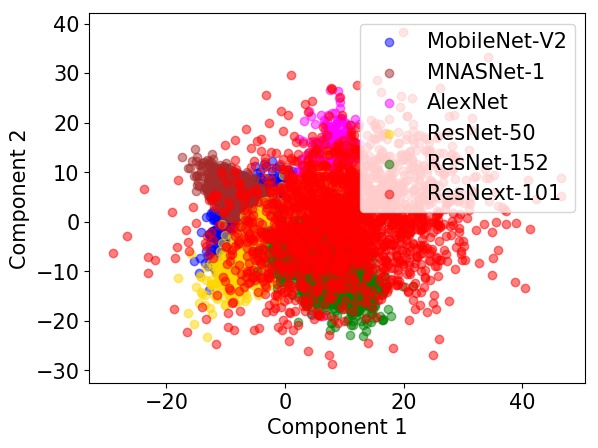}}
    \end{subfigure}
     \caption{The t-SNE visualization of feature space of validation set of ImageNet dataset for the benchmark models trained using the proposed loss function. Left: mobile-cloud collaborative inference using \textit{mobilenet\_v2} on the mobile side and \textit{resnext101\_32x8d} on the cloud side. Right: Ensemble of six benchmark CNNs which is suitable for cloud based intelligent services which host the replicas of the most-accurate model. For instance, instead of replicating \textit{resnext101\_32x8d} on six different servers, one can host these six CNNs plus the multiplexer which achieves less compute resource usage and higher accuracy. }
 \label{resulting_tsne}
\end{figure}

We demonstrate the effectiveness of the contrastive loss in Figure \ref{resulting_tsne}. The learned embedding space is similar to our target Venn diagram style depicted in Figure \ref{venn_loss}. The inputs which are only in the expertise domain of a certain model are pushed to the boundaries and the inputs which can be predicted correctly by multiple models are closer to the center. The separable embedding space that we create enables a light-weight neural multiplexer to effectively learn the multiplexing function.

\section{Conclusion and Future Work} \label{section.conclusion}
In this paper, we present an algorithm to multiplex a deep learning model to use depending on the input complexity and resource budgets. With the proposed algorithm, the mobile devices can host a small and mobile-friendly model and detect the inputs that are likely to be predicted correctly by local inference. Mobile devices will offload the inputs that they find hard to the cloud servers to be inferred by the larger models hosted in the cloud. The communication cost of the cloud-based inference dominates the local inference computation cost. As a result, it is desirable to offload as little as possible to the cloud and meet the accuracy requirements at the same time. Our results show that a user only needs to offload 32\% of the inputs to the cloud while achieving an accuracy even higher than the cloud-hosted model. Furthermore, the cloud providers offering APIs for cognitive tasks replicate their best-performing model in the server to called for any inputs regardless of their level of complexity. However, with this approach, they can host a wide range of small and large models and choose one depending on the input. It will save 2.85x of the cloud provider's compute resources while improving the accuracy by 4.55\% compared to deploying the most accurate model.
\bibliographystyle{IEEEtran}
\bibliography{references.bib}

\end{document}